\documentclass[12pt]{article}
\begin{document}
\begin{center}
{\bf SPONTANEOUS CP VIOLATION IN THE $U(3)$ NJL MODEL }\\
\vspace{5mm}
 S.I. Kruglov \\
\vspace{5mm}
\textit{International Educational Centre, 2727 Steeles Ave. W, \# 104, \\
Toronto, Ontario, Canada M3J 3G9}
\end{center}

\begin{abstract}
With the help of the functional integration method, the formation
of scalar, pseudoscalar condensates, and dynamical symmetry
breaking in the $U(3)$ four-fermion model has been investigated.
We show the possibility of spontaneous CP symmetry violation in
the model under consideration. The bosonization procedures of the
model are performed; the propagators of quarks, scalar and
pseudoscalar fields are calculated in one loop approximation. The
masses of the scalar and pseudoscalar mesons are evaluated.

\end{abstract}

\section{ Introduction}

It is possible nowadays to derive the effective quark-meson
Lagrangians from the fundamental quantum chromodynamics (QCD)
Lagrangian. The reformulation of QCD in terms of hadrons has not
been completed yet. Therefore, in the domain of low energy, some
phenomenological models are introduced. Local effective chiral
Lagrangians (ECL) \cite{Weinberg}, \cite{Gasser},
\cite{Buchvostov} can describe low energy physics of hadrons with
good accuracy. The instanton vacuum theory \cite{Shuryak} explains
the appearance of the chiral condensate which leads to the
dynamical symmetry breaking (DSB) and to the effective four-quark
interaction (for two flavours) \cite{Hooft} (see also
\cite{Bernard}, \cite{Kruglov90}). So, a contact four-fermion
interaction modelling quark interactions, takes into account both
quarks and mesons \cite{Eguchi}. In such models, the gluon
interactions are neglected and there is no confinement of quarks.
Therefore Nambu-Jona-Lasinio (NJL) models are QCD motivated
effective models with some shortcomings. In particular, NJL models
make it possible to decay the scalar mesons into $q\bar{q}$.

Our goal is to study the possibility of spontaneous CP symmetry
violation in the NJL model. The electric dipole moments of
particles violate CP-invariance and, in the framework of QCD, can
be explained with the help of the $\theta$-term. The effect of CP
breaking in strong interactions is small, but the investigation of
such phenomenon is important. It should be noted that the
$\theta$-term is important for the solution of the $U_A(1)$
problem. The axial symmetry, $U_A(1)$, is broken by the QCD
anomaly. This may be explained by the interactions of light quarks
and instantons which violate the $U_A(1)$ symmetry. There is a
region of quark masses \cite{Creutz}, where CP symmetry is
spontaneously broken. The CP violation leads to the exotic
phenomena, the possibility of $\eta$ decaying into two pions.

The present work is the generalization of \cite{Kruglov} on the
case of the U(3) group\footnote{Mass formulas were obtained in
\cite{Kruglov} on the basis of the specific relation (constraint)
between quadratic and logarithmic diverging integrals. In present
paper I use the standard cutoff regularization.}.

\section{Model and perturbation expansion}

We start with a NJL model possessing the internal symmetry group
$U(3)\otimes U(3)$ in the chiral limit:
\[
{\cal L }(x)=- \overline{\psi}(x)(\gamma_\mu \partial_\mu
+\widehat{m}_0)\psi (x)
\]
\vspace{-7mm}
\begin{equation} \label{1}
\end{equation}
\vspace{-7mm}
\[
+\frac{G}{2}\left\{\left[ \overline{\psi}(x)\lambda^a\psi
(x)\right]^2 + \left[ i\overline{\psi}(x)\gamma_5\lambda^a\psi
(x)\right]^2\right\} ,
\]
where $\lambda^a$ ($a=0,1,...,8$) are the Gell-Mann matrices,
$\lambda_0=\sqrt{2/3}I_8$ ($I_8$ is the unit $8\times8$-matrix),
$\partial_\mu =(\partial/\partial x_i ,-i\partial/\partial x_0)$
($x_0$ is the time), $\gamma_\mu$ are the Dirac matrices,
$\gamma_5=\gamma_1 \gamma_2 \gamma_3\gamma_4$. The $\widehat{m}_0$
is the matrix of bare masses of the quark triplet $\psi (x)$:
\[
\psi (x)=\mbox{diag}\left[u(x),d(x),s(x)\right] ,\hspace{0.3in}
\widehat{m}_0=\mbox{diag}\left(m_{u}, m_{d}, m_{s}\right) .
\]
The summation over colour quark degrees of freedom $n=1,2,...,N_C$
is implied here. The chiral symmetry is broken by quark masses and
dynamically by the appearance of condensates. Therefore, for
simplicity, we consider only the formation of the nonet of scalar
mesons and the nonet of pseudoscalar mesons $\pi$, $K$, $\eta$,
$\eta'$. The octets of vector and pseudovector mesons are ignored
here. The $U_A (1)$ symmetry is not broken here as the Lagrangian
(1) at $\widehat{m}_0=0$ is invariant under $\gamma_5$-chiral
transformations. To violate this symmetry ``by hands'', one can
add to the Lagrangian (1) the six-quark interaction due to
instantons \cite{Bernard}. On the other hand the $U_A (1)$ anomaly
appears because of the non-invariance of the fermion measure in
the functional integral \cite{Fujikawa}. It should be noted that
the QCD anomaly, $U_A (1)$, results the existence of a ninth
Goldstone boson $\eta'$ with the greater mass compared to $\eta$.

Using the functional integration method \cite{Pervushin}, the
generating functional for Green's functions
\begin{equation}
Z[ \overline{\eta},\eta]=N_0\int D \overline{\psi}D\psi
\exp\left\{i\int d^4 x\left[{\cal L}(x)+ \overline{\psi}(x)\eta
(x)+ \overline{\eta}(x)\psi (x)\right]\right\} , \label{2}
\end{equation}
where $ \overline{\eta}$, $\eta$ are external sources, with the
help of the replacement
\[
N_0=N\int D\phi_a D\widetilde{\phi}_a \exp \biggl\{
-i\frac{M^2}{2}\int d^4 x \biggl[\left(\widetilde{\phi}_a
(x)-i\frac{g_0}{M^2} \overline{\psi}(x)\gamma_5\lambda^a \psi
(x)\right)^2
\]
\[
+\left(\phi_a (x)-\frac{g_0}{M^2} \overline{\psi} (x)\lambda^a\psi
(x)\right)^2 \biggr]\biggr\} ,
\]
can be cast into
\[
Z[\overline{\eta},\eta]=N\int D \overline{\psi}D\psi D\phi_a
D\widetilde{\phi}_a \exp\biggl\{i\int d^4
x\biggl[-\overline{\psi}^n(x)\biggl[\gamma_\mu
\partial_\mu +\widehat{m}_0 -g_0\biggl( \phi_a (x)
\]
\vspace{-7mm}
\begin{equation} \label{3}
\end{equation}
\vspace{-7mm}
\[
+i\gamma_5 \widetilde{\phi}_a (x)\biggr)\lambda^a\biggr]\psi
(x)-\frac{M^2}{2}\left(\widetilde{\phi}_a^2 (x) + \phi_a^2
(x)\right) +\overline{\psi}(x)\eta (x) +\overline{\eta}(x)\psi (x)
\biggr]\biggr\} ,
\]
where $G=g_0^2/M^2$; $g_0$ is the dimensionless bare coupling
constant and $M$ is dimensional constant. Eq. (3) may be
integrated over the $\overline{\psi}$, $\psi$, and as a result
\begin{equation}
Z[\overline{\eta},\eta]=N\int D\phi_a D\widetilde{\phi}_a
\exp\biggl\{iS[\Phi] + i\int d^4 x~d^4 y~ \overline{\eta}(x)S_f
(x,y)\eta(y)\biggr\} , \label{4}
\end{equation}
where the effective action for bosonic collective fields $\Phi_a
(x)=\phi_a (x)+i\gamma_5 \widetilde{\phi}_a (x)$ is given by
\begin{equation}
S[\Phi]=-\frac{M^2}{2}\int d^4 x\left[\phi_a^2 (x)+
\widetilde{\phi}_a^2 (x)\right] -i \mbox{Tr} \ln \left[-\gamma_\mu
\partial_\mu -\widehat{m}_0 + g_0\Phi_a (x)\lambda^a\right] .
\label{5}
\end{equation}
We have used here the relation $\mbox{det} Q =
\mbox{exp}~\mbox{Tr}~\mbox{ln} Q$ (the $Q$ is an operator). The
operator $\mbox{Tr}$ in Eq. (5) includes tracing in matrix and
space-time variables. The Green function of quarks, $S_f (x,y)$,
obeys the equations
\begin{equation}
\left[\gamma_\mu \partial_\mu +\widehat{m}_0-g_0\Phi_a
(x)\lambda^a\right]S_f (x,y)=\delta(x-y) .
 \label{6}
\end{equation}
The fields $\phi_a (x)$ and $\widetilde{\phi}_a (x)$ form nonets
of scalar and pseudoscalar ($\pi^{\pm}$, $\pi^0$, $K^{\pm}$,
$K^0$, $\overline{K}^0$, $\eta$, $\eta'$) mesons.

The symmetric vacuum in the NJL models is not stable
\cite{Eguchi}. The physical vacuum is reconstructed which results
in the appearance of condensates and dynamical breaking of the
$U(3)\otimes U(3)$ symmetry. We imply here that the condensates
are formed as follows:
\[
\langle \overline{\psi}\psi\rangle\neq 0
,~~~\langle\overline{\psi}\lambda^3\psi\rangle\neq 0
,~~~\langle\overline{\psi}\lambda^8\psi\rangle\neq 0 ,
\]
\vspace{-8mm}
\begin{equation}
\label{7}
\end{equation}
\vspace{-8mm}
\[
 \langle \overline{\psi}\gamma_5\psi \rangle\neq 0
,~~~\langle\overline{\psi}\gamma_5\lambda^3\psi\rangle\neq 0
,~~~\langle\overline{\psi}\gamma_5\lambda^8\psi\rangle\neq 0 .
\]
The vacuum expectation values containing the $\gamma_5$ matrix are
parity and time reversal odd values, and as a result, they violate
CP symmetry. To take into consideration and to determine
condensates, the fields have to be ``shifted'' by the constants.
Therefore, we make the substitution in Eqs. (5), (6)
\[
\phi_0 (x)=\phi'_0 (x)+\sigma_0 , ~\phi_3 (x)=\phi'_3 (x)+\sigma_3
, ~\phi_8 (x)=\phi'_8 (x)+\sigma_8 , ~\phi_i (x)=\phi'_i (x) ,
\]
\vspace{-7mm}
\begin{equation}  \label{8}
\end{equation}
\vspace{-7mm}
\[
\widetilde{\phi}_0 (x)=\widetilde{\phi}'_0
(x)+\widetilde{\sigma}_0 , ~\widetilde{\phi}_3
(x)=\widetilde{\phi}'_3 (x)+\widetilde{\sigma}_3 ,
~\widetilde{\phi}_8 (x)=\widetilde{\phi}'_8
(x)+\widetilde{\sigma}_8 , ~\widetilde{\phi}_i
(x)=\widetilde{\phi}'_i (x) ,
\]
where $i=1,2,4,5,6,7$; $\sigma_0$, $\sigma_3$, $\sigma_8$,
$\widetilde{\sigma}_0$, $\widetilde{\sigma}_3$,
$\widetilde{\sigma}_8$ are coordinate -independent and
Lorentz-invariant constants. The fields $\phi'_a(x)$,
$\widetilde{\phi}'_a(x)$ in Eqs. (8) represent quantum excitations
over vacuum and are assumed to be small. The vacuum expectation
values (condensates), $\widetilde{\sigma}_0$,
$\widetilde{\sigma}_3$ and $\widetilde{\sigma}_8$, break CP
symmetry. Below, the condensates $\sigma_j$ and
$\widetilde{\sigma}_j$ for $j=0,3,8$ will be obtained from the
minimum of the effective potential defining the energy density of
the vacuum. To formulate the perturbation theory \cite{Pervushin},
we use the saddle-point method. Taking into consideration Eqs.
(8), one may rewrite Eq. (5) as follows:
\[
S[\Phi']=-\frac{M^2}{2}\int d^4 x\left[\left(\phi'_j
(x)+\sigma_j\right)^2+\left(\widetilde{\phi}'_j (x)+
\widetilde{\sigma}_j\right)^2+\phi_i^{'2}
+\widetilde{\phi}_i^{'2}\right]
\]
\vspace{-7mm}
\begin{equation}  \label{9}
\end{equation}
\vspace{-7mm}
\[
-i \mbox{Tr} \ln \left[-\gamma_\mu
\partial_\mu -\widehat{m}+i\widehat{\widetilde{m}}\gamma_5 + g_0\Phi'_a
(x)\lambda^a\right] ,
\]
where $\Phi'_a (x)=\phi'_a (x)+i\gamma_5 \widetilde{\phi}'_a (x)$,
$i=1,2,4,5,6,7$; $j=0,3,8$,
\[
\widehat{m}=\mbox{diag}\left(m_{01},m_{02},m_{03}\right)
,~~~\widehat{\widetilde{m}}=
\mbox{diag}\left(\widetilde{m}_1,\widetilde{m}_2,\widetilde{m}_3\right)
,
\]
\[
m_{01} =m_{u}-g_0\left(\sqrt{\frac{2}{3}}\sigma_0 +
\sigma_3+\frac{\sigma_8}{\sqrt{3}}\right) ,~ m_{02}
=m_{d}-g_0\left(\sqrt{\frac{2}{3}}\sigma_0
-\sigma_3+\frac{\sigma_8}{\sqrt{3}} \right) ,
\]
\vspace{-5mm}
\begin{equation}  \label{10}
\end{equation}
\vspace{-5mm}
\[
m_{03} =m_{s}-g_0\left(\sqrt{\frac{2}{3}}\sigma_0 -\frac{2
\sigma_8}{\sqrt{3}}\right) ,~~~
\widetilde{m}_1=g_0\left(\sqrt{\frac{2}{3}}\widetilde{\sigma}_0 +
\widetilde{\sigma}_3+\frac{\widetilde{\sigma}_8}{\sqrt{3}}\right)
,
\]
\[
\widetilde{m}_2=g_0\left(\sqrt{\frac{2}{3}}\widetilde{\sigma}_0
-\widetilde{\sigma}_3+\frac{\widetilde{\sigma}_8}{\sqrt{3}}
\right)
,~~~\widetilde{m}_3=g_0\left(\sqrt{\frac{2}{3}}\widetilde{\sigma}_0
-\frac{2 \widetilde{\sigma}_8}{\sqrt{3}}\right) .
\]
Let us consider the equality
\[
\mbox{Tr} \ln \left[-\gamma_\mu
\partial_\mu -\widehat{m}+i\widehat{\widetilde{m}}\gamma_5 +
g_0\Phi'_a (x)\lambda^a\right]
\]
\[
=\mbox{Tr} \ln \left(-\gamma_\mu
\partial_\mu -\widehat{m}
+i\widehat{\widetilde{m}}\gamma_5\right)+ \mbox{Tr} \ln
\left[1-g_0 S_{0f}(x,y)\Phi'_a (x)\lambda^a\right] ,
\]
where the Green function $S_{0f}(x,y)$ obeys the equation
\begin{equation}
\left[\gamma_\mu
\partial_\mu +\widehat{m}-i\widehat{\widetilde{m}}\gamma_5
\right]S_{0f}(x,y)=\delta(x-y) .
 \label{11}
\end{equation}
Then expanding the logarithm in small fluctuations $\Phi'_a (x)$,
the effective action (5) takes the form
\[
S[\Phi']=-\frac{M^2}{2}\int d^4 x\left[\left(\phi'_j
(x)+\sigma_j\right)^2+\left(\widetilde{\phi}'_j (x)+
\widetilde{\sigma}_j\right)^2+\phi_i^{'2}
+\widetilde{\phi}_i^{'2}\right]
\]
\vspace{-7mm}
\begin{equation}  \label{12}
\end{equation}
\vspace{-7mm}
\[
-i \mbox{Tr}\ln \left(-\gamma_\mu
\partial_\mu -\widehat{m}+i\widehat{\widetilde{m}}\gamma_5 \right)
+ \sum_{n=1}^{\infty}\frac{i}{n}\mbox{Tr}\left(g_0 S_{0f}\Phi_a '
\lambda^a\right)^n ,
\]
where
\[
\mbox{Tr}\left(g_0 S_{0f}\Phi_a ' \lambda^a\right)^n  =
\mbox{tr}\biggl[g_0^n\int~d^4x_1...d^4x_n~S_{0f}(x_n-x_1)\Phi_{a_1}
'\lambda^{a_1}
\]
\vspace{-7mm}
\begin{equation}  \label{13}
\end{equation}
\vspace{-7mm}
\[
\times S_{0f}(x_1-x_2)\Phi_{a_2}'\lambda^{a_2}...
S_{0f}(x_{n-1}-x_n)\Phi_{a_n}'\lambda^{a_n} \biggr] ,
\]
where the tr[...] means the tracing in matrices. The terms with
$n=2,3$ in Eq. (12) define decaying and the scattering of mesons.
The fields $\phi'_a$ and $\widetilde{\phi}'_a$ in the effective
action (12), after renormalization, describe the physical scalar
and pseudoscalar mesons.

\section{Propagators of quarks and mesons}

To calculate the masses of quarks and mesons it is necessary to
find the propagators of quarks and mesons. The condensates
$\sigma_j$ and $\widetilde{\sigma}_j$ for $j=0,3,8$ can be
obtained from the requirement that terms linear in fields
$\phi'_a(x)$, $\widetilde{\phi}'_a(x)$, which correspond to the
``tadpole'' diagrams, are absent in the effective action (12).
This leads to the gap equations
\[
\frac{\delta S[\Phi']}{\delta\phi'_j (x)}|_{\phi'_j
=0}=-M^2\sigma_j +ig_0\mbox{Tr}\left[S_{0f}(x,x)\lambda^j\right]=0
,
\]
\vspace{-7mm}
\begin{equation} \label{14}
\end{equation}
\vspace{-7mm}
\[
\frac{\delta S[\Phi']}{\delta\widetilde{\phi}'_j
(x)}|_{\widetilde{\phi}'=0}=-M^2\widetilde{\sigma}_j
(x)-g_0\mbox{Tr}\left[S_{0f}(x,x)\gamma_5\lambda^j\right]=0 .
\]
To find a solution of Eq. (11), we write down it in the momentum
space:
\begin{equation}
\left[i\widehat{p} +\widehat{m}-i\widehat{\widetilde{m}}\gamma_5
\right]S_{0f}(p)=1 ,
 \label{15}
\end{equation}
where $\widehat{p}=p_\mu \gamma_\mu$, $p_\mu=(\textbf{p},ip_0)$.
It is easy to verify that the solution to Eq. (15) for the Green
function is given by
\begin{equation}
S_{0f} (p)=\mbox{diag}\biggl(
\frac{-i\widehat{p}+m_{01}+i\widetilde{m}_1\gamma_5}{p^2 +m_1^2},
\frac{-i\widehat{p}+m_{02}+i\widetilde{m}_2\gamma_5}{p^2 +m_2^2},
\frac{-i\widehat{p}+m_{03}+i\widetilde{m}_3 \gamma_5}{p^2
+m_3^2}\biggr), \label{16}
\end{equation}
where
\begin{equation}
m_1^2=m_{01}^2+\widetilde{m}_1^2 ,~~
m_2^2=m_{02}^2+\widetilde{m}_2^2 ,~~
m_3^2=m^2_{03}+\widetilde{m}_3^2 . \label{17}
\end{equation}
The poles of the Green function (16) define the dynamical
(constituent) masses of $u$, $d$ and $s$ quarks: $m_1$, $m_2$,
$m_3$. The scalar ($\sigma_j$) and pseudoscalar
($\widetilde{\sigma}_j$) condensates contribute to the constituent
masses of all quarks. The terms containing $\widetilde{m}_j$ in
Eq. (16) violate CP symmetry. Substituting Eq. (16) into Eqs.
(14), one obtains a system of gap equations:
\[
M^2\sigma_0 =g_0\sqrt{\frac{2}{3}}\left(I_1m_{01} +I_2m_{02}
+I_3m_{03}\right) ,~~~ M^2\sigma_3 =g_0\left(I_1m_{01}
-I_2m_{02}\right) ,
\]
\[
M^2\sigma_8 =\frac{g_0}{\sqrt{3}}\left(I_1m_{01}
+I_2m_{02}-2I_3m_{03}\right) ,
\]
\begin{equation}
M^2\widetilde{\sigma}_0
=-g_0\sqrt{\frac{2}{3}}\left( \widetilde{m}_1 I_1 +\widetilde{m}_2
I_2 +\widetilde{m}_3 I_3\right) , \label{18}
\end{equation}
\[
M^2\widetilde{\sigma}_3 =-g_0\left( \widetilde{m}_1 I_1-
\widetilde{m}_2 I_2\right) ,~~~ M^2\widetilde{\sigma}_8
=-\frac{g_0}{\sqrt{3}}\left( \widetilde{m}_1 I_1 +\widetilde{m}_2
I_2 -2\widetilde{m}_3 I_3\right) ,
\]
where quadratic diverging integrals are given by
\begin{equation}
I_j =\frac{iN_C}{4\pi^4}\int \frac{d^4 p}{p^2
+m_j^2}=\frac{N_C}{4\pi^2}\left[m_j^2\ln\left(\frac{\Lambda^2}{m_j^2}
+1 \right)-\Lambda^2 \right] , \label{19}
\end{equation}
where $d^4 p=id^3 pdp_0 $, the $\Lambda$ is a cutoff and there is
no summation in index $j$ $(j=1,2,3)$ in Eq. (19). The
self-consistent equations (18) connect such parameters of a model
as the dimensional constant $G$, condensates $\sigma_j$ (or
dynamical masses of quarks) and a cutoff. The system of six gap
equations (18), defining the vacuum expectations $\sigma_j$,
$\widetilde{\sigma}_j$, with the help of Eqs. (10) can be
rewritten as
\begin{equation}
\left(m_{u} -m_{01}\right) =2m_{01}GI_1, \left(m_{d}
-m_{02}\right) =2m_{02}GI_2, \left(m_{s} -m_{03}\right)
=2m_{03}GI_3, \label{20}
\end{equation}
\begin{equation}
-\widetilde{m}_{1} =2\widetilde{m}_{1}GI_1 ,~~~~
-\widetilde{m}_{2} =2\widetilde{m}_{2}GI_2 ,~~~~
-\widetilde{m}_{3} =2\widetilde{m}_{3}GI_3 . \label{21}
\end{equation}
There are different solutions of gap equations (20), (21). We are
interested here in the possibilities of CP violation
($\widetilde{m}_j \neq 0$). Therefore consider the case when Eqs.
(21) have non-trivial solutions. It follows from Eqs. (21) that if
three vacuum expectations $\widetilde{m}_j $ ($j=1,2,3$) do not
equal zero, $\widetilde{m}_j\neq 0$, then $I_1=I_2=I_3$, and
therefore $m_1=m_2=m_3$. This is not interesting case because the
strange quark $s$ is much heavier than the $u$ and $d$ quarks.
Another solution is $\widetilde{m}_3 =0$, $\widetilde{m}_1 \neq
0$, $\widetilde{m}_2 \neq 0$. Then from Eqs. (21), we arrive at
the case $m_1=m_2$, $\widetilde{m}_1 =\widetilde{m}_2$, i.e.
isotopic symmetry is not broken, and the gap equation becomes
$2g_0^2I_1=-M^2$ ($I_1=I_2$). Comparing this equation with Eqs.
(20), one makes a conclusion that $m_{u}=m_d =0$, i.e. the chiral
limit for the light quarks is realized. We expect that pions
($\pi^{\pm}$, $\pi^0$) will be massless Goldstone particles in
this case. Requiring $m_s \neq 0$, one arrives from Eqs. (20) to
two gap equations
\begin{equation}
\left(m_{s} -m_{03}\right) =2m_{03}GI_3 ,~~~~ -1=2GI_1 .
\label{22}
\end{equation}
At the same time, if there is no CP violation, $\widetilde{m}_1 =
\widetilde{m}_2 =\widetilde{m}_3 =0$, we can analyze the case $m_u
\neq 0$, $m_d \neq 0$, $m_s \neq 0$, and gap equations (20) are
valid (see \cite{Bernard}, \cite{Eguchi} for other studies). We
pay attention here in the case $\widetilde{m}_1 =\widetilde{m}_2
\neq 0$, $\widetilde{m}_3 =0$, $m_u =m_d =0$, which requires (see
Eqs. (10))
\begin{equation}
\sigma_3 =\widetilde{\sigma}_3 =0 , ~~~~\widetilde{\sigma}_0
=\sqrt{2}\widetilde{\sigma}_8 .\label{23}
\end{equation}
The independent parameters here are the current quark mass
$m_{s}$, the cutoff $\Lambda$, and the dimensional constant $G$.

From Eq. (12), one may obtain the part of effective action which
does not depend on coordinates:
\begin{equation}
S[\sigma, \widetilde{\sigma}]=-\frac{M^2}{2}\int d^4 x\left[\left(
\sigma_j\right)^2+\left(\widetilde{\sigma}_j\right)^2 \right] -i
\mbox{Tr}\ln \left(-\gamma_\mu\partial_\mu - \widehat{m}
+i\widehat{\widetilde{m}}\gamma_5 \right) . \label{24}
\end{equation}
We can use the relation $S[\sigma, \widetilde{\sigma}]=-\int d^4
x~V_{eff}$ \cite{Jona} for the constant fields. As a result, one
may get from Eq. (24), with the help of Eq. (10), the effective
potential
\[
V_{eff}=\frac{M^2}{4g_0^2}\left[\left(
m_{01}-m_{u}\right)^2+\left(m_{02}-m_{d}\right)^2 +
\left(m_{03}-m_{s}\right)^2 \right]
\]
\vspace{-7mm}
\begin{equation} \label{25}
\end{equation}
\vspace{-7mm}
\[
+ \frac{i N_C}{8\pi^4}\int d^4 p~\ln \left(p^2 +m_1^2 \right)
\left(p^2 +m_2^2 \right)\left(p^2 +m_3^2 \right) .
\]
We will keep all parameters to be nonzero for the possibility to
study as the case with CP violation as well as the case without CP
breaking. Eqs. (18) or (20), (21) may be obtained from the
condition of effective potential (25) to realize the minimum:
\begin{equation}
\frac{\partial V_{eff}}{\partial m_{0j}}=\frac{\partial
V_{eff}}{\partial \widetilde{m}_j}=0 ~~~~(j=1,2,3) . \label{26}
\end{equation}
To obtain the mass spectrum of mesons, one needs to evaluate the
terms in Eq. (12), quadratic in fields $\phi'_a$,
$\widetilde{\phi}'_a$. From Eq. (12) we find
\[
S^{(2)}[\Phi']=-\frac{M^2}{2}\int d^4 x\left[\phi_a^{'2}
+\widetilde{\phi}_a^{'2}\right]+ \frac{i}{2}\mbox{Tr}\left(g_0
S_{0f}\Phi_a '\lambda^a\right)^2
\]
\vspace{-7mm}
\begin{equation} \label{27}
\end{equation}
\vspace{-7mm}
\[
\equiv -\frac{1}{2}\int d^4x~d^4y~\phi'_A (x)\Delta^{-1}_{AB}
(x,y)\phi'_B (y) .
\]
In the momentum space the inverse meson symmetric propagator is
given by
\begin{equation}
\Delta^{-1}_{AB}(p)=-ig_0^2 \mbox{tr}\left[\int\frac{d^4
k}{(2\pi^4)} S_{0f} (k)T_A S_{0f} (k-p)T_B\right]+\delta_{AB}M^2_A
, \label{28}
\end{equation}
where $T_A=(\lambda^a,i\gamma_5\lambda^a)$, $\phi_A '=(\phi_a
',\widetilde{\phi}_a ')$, and we use the notation
$A=(a,\widetilde{a})$.

Evaluating the traces in Eqs. (28), we obtain the nonzero elements
of the inverse propagators of the scalar ($\Phi_a'(x)$) mesons:
\[ \Delta_{00}^{-1}(p)=M^2+g_0^2\frac{2}{3}\left( I_1 +I_2
+I_3\right)+\frac{1}{3}\left(p^2+4 m_{01}^2\right)I_{11}(p)
\]
\[
+\frac{1}{3}\left(p^2+4 m_{02}^2\right)I_{22}(p)
+\frac{1}{3}\left(p^2+4 m_{03}^2\right)I_{33}(p) ,
\]
\[
\Delta_{11}^{-1}(p)=\Delta_{22}^{-1}(p)
=M^2+g_0^2\left(I_1+I_2\right) +\left[p^2
+\left(m_{02}+m_{01}\right)^2
+\left(\widetilde{m}_{1}-\widetilde{m}_{2}\right)^2\right]I_{12}(p)
,
\]
\[
\Delta_{33}^{-1}(p) =M^2+g_0^2\left( I_1+I_2\right)
+\frac{1}{2}\left(p^2 +4m_{01}^2 \right)I_{11}(p) +
\frac{1}{2}\left(p^2 +4m_{02}^2 \right)I_{22}(p) ,
\]
\[
\Delta_{44}^{-1}(p)=\Delta_{55}^{-1}(p)
=M^2+g_0^2\left(I_1+I_3\right) +\left[p^2
+\left(m_{03}+m_{01}\right)^2
+\left(\widetilde{m}_{1}-\widetilde{m}_{3}\right)^2\right]I_{13}(p)
,
\]
\begin{equation}
\Delta_{66}^{-1}(p)=\Delta_{77}^{-1}(p)
=M^2+g_0^2\left(I_2+I_3\right) +\left[p^2
+\left(m_{03}+m_{02}\right)^2
+\left(\widetilde{m}_{2}-\widetilde{m}_{3}\right)^2\right]I_{23}(p)
, \label{29}
\end{equation}
\[
\Delta_{88}^{-1}(p)=M^2+\frac{g_0^2}{3}\left(I_1 +I_2
+4I_3\right)+\frac{1}{6}\left(p^2+4m_{01}^2\right)I_{11}(p)
\]
\[
+\frac{1}{6}\left(p^2+4m_{02}^2\right)I_{22}(p)+
\frac{2}{3}\left(p^2+4m_{03}^2\right)I_{33}(p),
\]
\[
\Delta_{03}^{-1}(p)=g_0^2\sqrt{\frac{2}{3}}\left(I_1 -I_2\right)
+\frac{1}{\sqrt{6}}\left(p^2+4m_{01}^2 \right)I_{11}(p) -
\frac{1}{\sqrt{6}}\left(p^2 +4m_{02}^2 \right)I_{22}(p) ,
\]
\[
\Delta_{08}^{-1}(p)=\frac{g_0^2 \sqrt{2}}{3} \left(I_1 +I_2 -2I_3
\right)+\frac{\sqrt{2}}{6}\left(p^2 +4m_{01}^2 \right)I_{11}(p)
\]
\[
+\frac{\sqrt{2}}{6}\left(p^2 +4m_{02}^2 \right)I_{22}(p)
-\frac{\sqrt{2}}{3}\left(p^2 +4m_{03}^2 \right)I_{33}(p)
,~~~~~\sqrt{2}\Delta_{38}^{-1}(p)=\Delta_{03}^{-1}(p) .
\]
One can get from Eq. (28) the inverse propagators of pseudoscalar
($\widetilde{\Phi}_a '(x)$) mesons:
\[
\Delta_{\tilde{0}\tilde{0}}^{-1}(p)=M^2+g_0^2\frac{2}{3}\left(I_1
+I_2 +I_3 \right)+\frac{1}{3}\left(p^2+4
\widetilde{m}_1^2\right)I_{11}(p)
\]
\[
+\frac{1}{3}\left(p^2+4 \widetilde{m}_2^2\right)I_{22}(p)
+\frac{1}{3}\left(p^2+4 \widetilde{m}_3^2\right)I_{33}(p) ,
\]
\[
\Delta_{\tilde{1}\tilde{1}}^{-1}(p)=\Delta_{\tilde{2}\tilde{2}}^{-1}(p)
=M^2+g_0^2\left(I_1+I_2\right)
\]
\[
+\left[p^2 +\left(m_{02}-m_{01}\right)^2
+\left(\widetilde{m}_{1}+\widetilde{m}_{2}\right)^2\right]I_{11}(p)
,
\]
\[
\Delta_{\tilde{3}\tilde{3}}^{-1}(p) =M^2+g_0^2\left(
I_1+I_2\right) +\frac{1}{2}\left(p^2 +4\widetilde{m}_1^2
\right)I_{11}(p) + \frac{1}{2}\left(p^2 +4\widetilde{m}_2^2
\right)I_{22}(p) ,
\]
\begin{equation}
\Delta_{\tilde{4}\tilde{4}}^{-1}(p)=\Delta_{\tilde{5}\tilde{5}}^{-1}(p)
=M^2+g_0^2\left(I_1+I_3\right) \label{30}
\end{equation}
\[
+\left[p^2 +\left(m_{03}-m_{01}\right)^2
+\left(\widetilde{m}_{1}+\widetilde{m}_{3}\right)^2\right]I_{13}(p)
,
\]
\[
\Delta_{\tilde{6}\tilde{6}}^{-1}(p)=\Delta_{\tilde{7}\tilde{7}}^{-1}(p)
=M^2+g_0^2\left(I_2+I_3\right)
\]
\[
+\left[p^2 +\left(m_{03}-m_{02}\right)^2
+\left(\widetilde{m}_{2}+\widetilde{m}_{3}\right)^2\right]I_{23}(p)
,
\]
\[
\Delta_{\tilde{8}\tilde{8}}^{-1}(p)=M^2+\frac{g_0^2}{3}\left(I_1
+I_2+4I_3\right)+\frac{1}{6}\left(p^2+4\widetilde{m}_1^2\right)I_{11}(p)
\]
\[
+\frac{1}{6}\left(p^2+4\widetilde{m}_2^2\right)I_{22}(p)+
\frac{2}{3}\left(p^2+4\widetilde{m}_3^2\right)I_{33}(p) ,
\]
\[
\Delta_{\tilde{0}\tilde{3}}^{-1}(p)
=g_0^2\sqrt{\frac{2}{3}}\left(I_1
-I_2\right)+\frac{1}{\sqrt{6}}\left(p^2 +4\widetilde{m}_1^2
\right)I_{11}(p) - \frac{1}{\sqrt{6}}\left(p^2 +4\widetilde{m}_2^2
\right)I_{22}(p) ,
\]
\[
\Delta_{\tilde{0}\tilde{8}}^{-1}(p)=\frac{g_0^2 \sqrt{2}}{3}
\left(I_1 +I_2 -2I_3 \right)+\frac{\sqrt{2}}{6}\left(p^2
+4\widetilde{m}_1^2 \right)I_{11}(p)
\]
\[
+\frac{\sqrt{2}}{6}\left(p^2 +4\widetilde{m}_2^2 \right)I_{22}(p)
-\frac{\sqrt{2}}{3}\left(p^2 +4\widetilde{m}_3^2 \right)I_{33}(p)
,~~~~~\Delta_{\tilde{0}\tilde{3}}^{-1}(p)=\sqrt{2}\Delta_{\tilde{3}\tilde{8}}^{-1}(p)
.
\]
Non-diagonal scalar-pseudoscalar elements of inverse propagators
are given by
\[
\Delta_{\tilde{0}0}^{-1}(p)= - \frac{4}{3}
\left[m_{01}\widetilde{m}_1 I_{11}(p) +m_{02}\widetilde{m}_2
I_{22}(p) +m_{03} \widetilde{m}_3 I_{33}(p)\right] ,
\]
\[
\Delta_{8\tilde{0}}^{-1}(p)=\Delta_{0\tilde{8}}^{-1}(p)=\frac{2\sqrt{2}}{3}
\left[ 2m_{03}\widetilde{m}_3 I_{33}(p)-m_{01}\widetilde{m}_1
I_{11}(p) -m_{02}\widetilde{m}_2 I_{22}(p)\right] ,
\]
\vspace{-7mm}
\begin{equation}
\label{31}
\end{equation}
\vspace{-7mm}
\[
\Delta_{3\tilde{0}}^{-1}(p)=\Delta_{0\tilde{3}}^{-1}(p)=
\sqrt{2}\Delta_{8\tilde{3}}^{-1}(p)=\sqrt{2}\Delta_{3\tilde{8}}^{-1}(p)
\]
\[
= 2\sqrt{\frac{2}{3}}\left[m_{02}\widetilde{m}_2 I_{22}(p)- m_{01}
\widetilde{m}_1 I_{11}(p)\right] ,
\]
where the quadratic diverging integrals read
\[
I_{ij}(p)=-\frac{ig_0^2 N_C}{4\pi^4}\int \frac{d^4
k}{\left(k^2+m^2_i\right)\left[\left(k-p\right)^2+m^2_j\right]}
\]
\vspace{-7mm}
\begin{equation}
\label{32}
\end{equation}
\vspace{-7mm}
\[
=\frac{g_0^2N_C}{4\pi^2}\left[\ln\left(\frac{\Lambda^2}{m_i^2}\right)
-1 -\int_0^1 dx~\ln\frac{m_j^2+x\left(m_i^2 -m_j^2\right)+p^2
x(1-x)}{m_i^2}\right] ,
\]
and there is no summation in indexes $i,j$. Inverse propagators
(29)-(31) define spectrum of mass for the general case including
CP violation.

\section{Effective action and mass spectrum \\
of mesons}

Poles of the propagators (28) give the masses of mesons which can
be estimated by numerical calculations. Here we make some
evaluations of meson masses. From Eqs. (29)-(31), one can find the
effective action of the mesonic ``free'' fields
\begin{equation}
S_{free}[\Phi] =-\frac{1}{2}\int d^4 x\left[\left(\partial_\mu
\Phi_A (x)\right)^2+m^2_{AB}\phi_A (x)\phi_B (x)\right] ,
\label{33}
\end{equation}
where $A=(a,\tilde{a})$, $\phi_{\tilde{a}}\equiv
\widetilde{\phi}_a$. The eigenvalues of the symmetric mass matrix
$m^2_{AB}$ define the mass spectrum. To obtain the mass matrix, we
renormalize fields $\widetilde{\phi}_a
(x)=Z^{-1/2}\widetilde{\phi}_a '(x)$, $\phi_a (x)=Z^{-1/2}\phi_a
'(x)$, and the constant $g^2=Z g_0^2$, so that the variables
$g\phi_a$, $g\widetilde{\phi}_a$ are the renormalization-invariant
values. It follows from Eq. (32) that the renormalization constant
can be defined as follows
\begin{equation}
Z^{-1} =\frac{g_0^2 N_C}{4\pi^2}\left[ \ln \left(
\frac{\Lambda^2}{m_1^2}\right) -1\right] . \label{34}
\end{equation}
It is seen from Eq. (34) that the expansion in $g^2/4\pi^2$,
corresponds to the $1/N_C$ expansion. We imply here a cutoff
$\Lambda$ is chosen in such a way that $g^2/4\pi^2<1$. Using the
gap equations (20), (21), in the leading order, we find from Eqs.
(29) the elements of the mass matrix for scalar mesons:
\[
m_{00}^2=g^2\frac{2}{3}\left( \frac{m_u I_1}{m_u -m_{01}}
+\frac{m_d I_2}{m_d -m_{02}}+\frac{m_s I_3}{m_s
-m_{03}}\right)+\frac{4}{3}\left( m_{01}^2 + m_{02}^2
+m_{03}^2\right) ,
\]
\[
m_{11}^{2}=m_{22}^{2} = g^2\left( \frac{m_u I_1}{m_u -m_{01}}
+\frac{m_d I_2}{m_d -m_{02}}\right)+\left(m_{02}+m_{01}\right)^2
+\left(\widetilde{m}_{1}-\widetilde{m}_{2}\right)^2 ,
\]
\[
m_{33}^{2} =g^2\left( \frac{m_u I_1}{m_u -m_{01}} +\frac{m_d
I_2}{m_d -m_{02}}\right) +2\left(m_{01}^2 +m_{02}^2\right) ,
\]
\[
m_{44}^{2}=m_{55}^{2} = g^2\left( \frac{m_u I_1}{m_u -m_{01}}
+\frac{m_s I_3}{m_s -m_{03}}\right)+\left(m_{03}+m_{01}\right)^2
+\left(\widetilde{m}_{1}-\widetilde{m}_{3}\right)^2 ,
\]
\begin{equation}
m_{66}^{2}=m_{77}^{2} =g^2\left( \frac{m_d I_2}{m_d -m_{02}}
+\frac{m_s I_3}{m_s -m_{03}}\right) +\left(m_{03}+m_{02}\right)^2
+\left(\widetilde{m}_{2}-\widetilde{m}_{3}\right)^2 , \label{35}
\end{equation}
\[
m_{88}^{2}=\frac{g^2}{3}\left( \frac{m_u I_1}{m_u -m_{01}}
+\frac{m_d I_2}{m_d -m_{02}}+\frac{4m_s I_3}{m_s
-m_{03}}\right)+\frac{2}{3}\left( m_{01}^2 +m_{02}^2 +4m_{03}^2
\right) ,
\]
\[
m_{08}^{2}=\frac{g^2 \sqrt{2}}{3} \left(I_1 +I_2 -2I_3
\right)+\frac{2 \sqrt{2}}{3}\left(m_{01}^2 +m_{02}^2
-2m_{03}^2\right) ,
\]
\[
m_{03}^{2}=\sqrt{2}m_{38}^{2}=g^2\sqrt{\frac{2}{3}}\left(I_1
-I_2\right)+2\sqrt{\frac{2}{3}}\left(m_{01}^2 -m_{02}^2\right) .
\]
One can obtain from Eqs. (30) the elements of the mass matrix for
pseudoscalar mesons:
\[
m_{\tilde{0}\tilde{0}}^{2}=g^2\frac{2}{3}\left( \frac{m_u I_1}{m_u
-m_{01}} +\frac{m_d I_2}{m_d -m_{02}}+\frac{m_s I_3}{m_s
-m_{03}}\right) + \frac{4}{3}\left(\widetilde{m}_1^2
+\widetilde{m}_2^2 +\widetilde{m}_3^2\right) ,
\]
\[
m_{\tilde{0}\tilde{8}}^{2}=\frac{g^2 \sqrt{2}}{3} \left(I_1 +I_2
-2I_3 \right)+\frac{2 \sqrt{2}}{3}\left(\widetilde{m}_1^2
+\widetilde{m}_2^2 -2\widetilde{m}_3^2\right) ,
\]
\[
m_{\tilde{0}\tilde{3}}^{2}=\sqrt{2}m_{\tilde{3}\tilde{8}}^{2}=g^2\sqrt{\frac{2}{3}}\left(I_1
-I_2\right)+2\sqrt{\frac{2}{3}}\left(\widetilde{m}_1^2
-\widetilde{m}_2^2\right) ,
\]
\[
m_{\tilde{1}\tilde{1}}^{2}=m_{\tilde{2}\tilde{2}}^{2} =g^2\left(
\frac{m_u I_1}{m_u -m_{01}} +\frac{m_d I_2}{m_d
-m_{02}}\right)+\left(m_{02}-m_{01}\right)^2
+\left(\widetilde{m}_{1}+\widetilde{m}_{2}\right)^2,
\]
\begin{equation}
m_{\tilde{4}\tilde{4}}^{2}=m_{\tilde{5}\tilde{5}}^{2} = g^2\left(
\frac{m_u I_1}{m_u -m_{01}} +\frac{m_s I_3}{m_s
-m_{03}}\right)+\left(m_{03}-m_{01}\right)^2
+\left(\widetilde{m}_{1}+\widetilde{m}_{3}\right)^2 ,\label{36}
\end{equation}
\[
m_{\tilde{6}\tilde{6}}^{2}=m_{\tilde{7}\tilde{7}}^{2} =g^2\left(
\frac{m_d I_2}{m_d -m_{02}} +\frac{m_s I_3}{m_s -m_{03}}\right)
+\left(m_{03}-m_{02}\right)^2
+\left(\widetilde{m}_{2}+\widetilde{m}_{3}\right)^2 ,
\]
\[
m_{\tilde{3}\tilde{3}}^{2} =g^2\left( \frac{m_u I_1}{m_u -m_{01}}
+\frac{m_d I_2}{m_d -m_{02}}\right)+2\left(\widetilde{m}_1^2
+\widetilde{m}_2^2\right) ,
\]
\[
m_{\tilde{8}\tilde{8}}^{2}=\frac{g^2}{3}\left( \frac{m_u I_1}{m_u
-m_{01}} +\frac{m_d I_2}{m_d -m_{02}}+\frac{4m_s I_3}{m_s
-m_{03}}\right)+\frac{2}{3}\left( \widetilde{m}_1^2
+\widetilde{m}_2^2 +4\widetilde{m}_3^2 \right)^2 .
\]

Using Eqs. (31), we find non-diagonal scalar-pseudoscalar elements
of the mass matrix
\[
m_{\tilde{0}0}^{2}= - \frac{4}{3} \left(m_{01}\widetilde{m}_1
+m_{02}\widetilde{m}_2 +m_{03} \widetilde{m}_3 \right) ,
\]
\begin{equation}
m_{8\tilde{0}}^{2}=m_{0\tilde{8}}^{2}=\frac{2\sqrt{2}}{3} \left(
2m_{03}\widetilde{m}_3 -m_{01}\widetilde{m}_1
-m_{02}\widetilde{m}_2 \right)  , \label{37}
\end{equation}
\[
m_{3\tilde{0}}^{2}=m_{0\tilde{3}}^{2}=
\sqrt{2}m_{8\tilde{3}}^{2}=\sqrt{2}m_{3\tilde{8}}^{2}=
2\sqrt{\frac{2}{3}}\left(m_{02}\widetilde{m}_2 - m_{01}
\widetilde{m}_1 \right) .
\]
It is seen from Eqs. (35)-(37) the Goldstone nature of
pseudoscalar mesons: if bare masses of quarks are zero,
$\sigma_0=\sigma_3=\sigma_8=0$, $\widetilde{m}_j=0$, all
pseudoscalar mesons are massless. We recall that if
$\widetilde{m}_j \neq 0$ ($j=1,2$), gap equations require the
chiral limit: $m_u =m_d =0$. If there is no CP violation
($\widetilde{m}_j = 0$), one can consider the case $m_u \neq 0$,
$m_d \neq 0$ to have nonzero pion masses. It follows from Eq. (37)
that there is mixing of scalar $\phi_a (x)$ and pseudoscalar
$\widetilde{\phi}_a (x)$ fields due to the CP-violating
condensates $\widetilde{m}_j$. The fields $\widetilde{\phi}_0$ and
$\widetilde{\phi}_8$ are also mixed corresponding to $\eta -\eta'$
mixing. Pions, connected with the fields $\widetilde{\phi}_i (x)$
($i=1,2,3$), acquire nonzero masses due to the presence of the
CP-violating condensates even for zero current masses $m_u =m_d =
0$. At the same time in the case $\widetilde{m}_1 =
\widetilde{m}_2$, $\widetilde{m}_3 = 0$ there is less contribution
of CP violating condensates to the masses of scalar mesons.

To obtain the diagonal matrix $m_{AB}$, one can make the
transformation of the rotation group for fields $\phi_a (x)$,
$\widetilde{\phi}_a (x)$ ($a=0,3,8$). For the simple mixing of
fields $\widetilde{\phi}_0 (x)$, $\widetilde{\phi}_8 (x)$, one
obtains
\[
\widetilde{\phi}'_0 (x)=\widetilde{\phi}_0 (x)\cos\theta_P
-\widetilde{\phi}_8 (x)\sin\theta_P ,
\]
\vspace{-7mm}
\begin{equation}
\label{38}
\end{equation}
\vspace{-7mm}
\[
\widetilde{\phi}'_8 (x)=\widetilde{\phi}_0 (x)\sin\theta_P
+\widetilde{\phi}_8 (x)\cos\theta_P ,
\]
where
$\tan2\theta_P=2m_{\tilde{0}\tilde{8}}^2/(m_{\tilde{8}\tilde{8}}^2-m_{\tilde{0}\tilde{0}}^2)$.
The masses of bosonic fields $\widetilde{\phi}'_0 (x)$,
$\widetilde{\phi}'_8 (x)$ became:
\[
m_{\tilde{0}\tilde{0}}'^2=m_{\tilde{0}\tilde{0}}^2\cos^2
\theta_P+m_{\tilde{8}\tilde{8}}^2 \sin^2 \theta_P
-m_{\tilde{0}\tilde{8}}^2\sin 2\theta_P ,
\]
\vspace{-8mm}
\begin{equation}
\label{39}
\end{equation}
\vspace{-8mm}
\[
m_{\tilde{8}\tilde{8}}'^2=m_{\tilde{0}\tilde{0}}^2\sin^2
\theta_P+m_{\tilde{8}\tilde{8}}^2 \cos^2 \theta_P
+m_{\tilde{0}\tilde{8}}^2 \sin 2\theta_P .
\]

We consider the case $m_1=m_2$ when the isotopic symmetry is
conserved. It follows then from Eqs. (10) that this requires the
vacuum expectation value $\sigma_3 =0$. The relation
$m_1-m_{u}=-g(\sqrt{2}\sigma_0 +\sigma_8)/\sqrt{3}$ (after the
renormalization of the constant $g_0$) is treated as the quark
level version of the Goldberger-Treiman identity \cite{Goldberger}
with the pion decay constant $(\sqrt{2}\sigma_0
+\sigma_8)/\sqrt{3}=f_\pi =93$ MeV. We imply here a very small
possible contribution of CP violating condensates to the real
masses of mesons. Putting here the value of the constant
\cite{Scadron} $g=3.628$, one finds that the parameter of
expansion is $g^2/4\pi^2=1/N_C=1/3$. Using the freedom in the
choice of the bare quark mass, we set $m_u=m_d=5.3$ MeV. From the
Goldberger-Treiman relation one obtains the constituent masses of
the light quarks $m_1=m_2=342.7$ MeV ($\widetilde{m}_j=0$). It
follows from Eq. (34): the covariant cutoff $\Lambda$ is given by
$\Lambda=e m_1=931.5$ MeV. To find the constituent mass of the
s-quark, we find from the gap equations (20) the self-consistent
relation $m_1(m_s-m_3)I_1 = m_3(m_u-m_1)I_3$. Setting the free
parameter of the s-quark current mass $m_s=166$ MeV, for a given
cutoff, one obtains the dynamical strange quark mass: $m_3=570$
MeV. With the help of these masses and the cutoff $\Lambda$, we
calculate from Eqs. (36) the masses of $\pi$, $K$ mesons and quark
condensates
\[
m_{\pi}=139~\mbox{MeV} ,~~~~m_{K}=494~\mbox{MeV} ,
\]
\vspace{-7mm}
\begin{equation}
\label{40}
\end{equation}
\vspace{-7mm}
\[
\langle\overline{u}u\rangle =\langle\overline{d}d\rangle
=m_{1}I_1=\left(-252~\mbox{MeV}\right)^3
,~~~~\langle\overline{s}s\rangle
=m_{3}I_3=\left(-268~\mbox{MeV}\right)^3 .
\]
Masses of $K$ mesons are degenerated here as well as masses of
pions. The masses and condensates (40) are agreed with the
phenomenology. The pseudoscalar $\eta' - \eta$ mixing angle,
evaluated from Eqs. (38), is $\theta_P=-35^\circ$. Masses of
$\eta$, $\eta'$ and their mixing angle are not described correctly
here because we did not take into consideration anomaly and the
axial symmetry $U_A (1)$ is not broken. It is easy to verify that
the Gell-Mann--Oakes--Renner \cite{Gell} relation $f_\pi^2
m_\pi^2=-2m_u \langle\overline{u}u \rangle$ is approximately
valid.

From Eqs. (35) we obtain the elements of the mass matrix
corresponding to the nonet of scalar mesons
\[
m_{00}=938~\mbox{MeV} ,~~~~m_{11}=m_{22}=m_{33}=699~\mbox{MeV} ,
\]
\vspace{-8mm}
\begin{equation}
\label{41}
\end{equation}
\vspace{-8mm}
\[
m_{44}=m_{55}=m_{66}=m_{77}=1013~\mbox{MeV}
,~~~~m_{88}=1128~\mbox{MeV} .
\]
The mixing angle of the $\phi_0$ and $\phi_8$ fields is
$\theta_S=-35^\circ$. Scalar and pseudoscalar fields are not mixed
in the case (see Eqs. (37)) when the equality $\widetilde{m}_j =0$
is valid. We do not identify here the scalar mesons $\phi_a (x)$
with the nonet of known scalar mesons: $\sigma (560)$, $f_0(980)$,
$\kappa (900)$, $a_0(980)$ due to their complicated nature: there
are contributions of four-quark states and gluons in these mesons
\cite{Jaffe}, \cite{Black}, \cite{Achasov}.

\section{Conclusion}

The model under consideration can describe CP violation in strong
interactions. There is a contribution of CP violating condensates,
$\widetilde{m}_j$, to constituent masses of u, d and s quarks and
to masses of scalar and pseudoscalar mesons. If the current masses
of quarks equal zero, and the CP-violating condensate
$\widetilde{m}_j=0$, all pseudoscalar mesons $\pi$, $K$, $\eta$,
$\eta '$ become massless Goldstone bosons. Masses of all K-mesons
are degenerated in the case $m_{01}=m_{02}$. In this model, the
appearance of CP-violating condensates leads to the chiral limit:
$m_u=m_d=0$ . In the particular case $\widetilde{m}_j=0$, when
there is no CP violation, the model gives reasonable dynamical
quark masses, masses of $\pi$, $K$ mesons and quark condensates.
At the same time $\eta$ and $\eta'$ mesons can not be described
correctly in the framework of the model considered because the
$U_A(1)$ symmetry is not broken. To take into consideration the
$U_A(1)$-anomaly, one may generalize the model by including the
determinant six-quark interaction (due to instantons) violating
$U_A(1)$ symmetry \cite{Bernard}.

\section{Acknowledgments}

I wish to thank Prof. R. L. Jaffe for useful discussions.

\end{document}